\documentclass[aps,prl,preprint,superscriptaddress,endfloats*]{revtex4-1}

\usepackage[pdfborder={0 0 0},colorlinks=true,allcolors=blue]{hyperref}
\usepackage{graphicx}
\usepackage{xcolor}

\begin{document}


\title{Three-Dimensional Shapes of Spinning Helium Nanodroplets}



\author{Bruno Langbehn}
\email[]{bruno.langbehn@physik.tu-berlin.de}
\affiliation{Institut f\"ur Optik und Atomare Physik, Technische Universit\"at Berlin, 10623 Berlin, Germany}

\author{Katharina Sander}
\affiliation{Institut f\"ur Physik, Universit\"at Rostock, 18051 Rostock, Germany}

\author{Yevheniy Ovcharenko}
\affiliation{Institut f\"ur Optik und Atomare Physik, Technische Universit\"at Berlin, 10623 Berlin, Germany}
\affiliation{European XFEL GmbH, 22869 Schenefeld, Germany}

\author{Christian Peltz}
\affiliation{Institut f\"ur Physik, Universit\"at Rostock, 18051 Rostock, Germany}

\author{Andrew Clark}
\affiliation{Laboratory of Molecular Nanodynamics, Ecole Polytechnique F\'ed\'erale de Lausanne (EPFL), 1015 Lausanne, Switzerland}

\author{Marcello Coreno}
\affiliation{ISM-CNR, Istituto di Struttura della Materia, LD2 Unit, 34149 Trieste, Italy}

\author{Riccardo Cucini}
\affiliation{Elettra -- Sincrotrone Trieste S.C.p.A., 34149 Trieste, Italy}

\author{Marcel Drabbels}
\affiliation{Laboratory of Molecular Nanodynamics, Ecole Polytechnique F\'ed\'erale de Lausanne (EPFL), 1015 Lausanne, Switzerland}

\author{Paola Finetti}
\affiliation{Elettra -- Sincrotrone Trieste S.C.p.A., 34149 Trieste, Italy}

\author{Michele Di Fraia}
\affiliation{Elettra -- Sincrotrone Trieste S.C.p.A., 34149 Trieste, Italy}
\affiliation{ISM-CNR, Istituto di Struttura della Materia, LD2 Unit, 34149 Trieste, Italy}

\author{Luca Giannessi}
\affiliation{Elettra -- Sincrotrone Trieste S.C.p.A., 34149 Trieste, Italy}

\author{Cesare Grazioli}
\affiliation{ISM-CNR, Istituto di Struttura della Materia, LD2 Unit, 34149 Trieste, Italy}

\author{Denys Iablonskyi}
\affiliation{Institute of Multidisciplinary Research for Advanced Materials, Tohoku University, Sendai 980-8577, Japan}

\author{Aaron C. LaForge}
\affiliation{Physikalisches Institut, Universit\"at Freiburg, 79104 Freiburg, Germany}

\author{Toshiyuki Nishiyama}
\affiliation{Division of Physics and Astronomy, Graduate School of Science, Kyoto University, Kyoto 606-8502, Japan}

\author{Ver\'onica Oliver \'Alvarez de Lara}
\affiliation{Laboratory of Molecular Nanodynamics, Ecole Polytechnique F\'ed\'erale de Lausanne (EPFL), 1015 Lausanne, Switzerland}

\author{Paolo Piseri}
\affiliation{CIMAINA and Dipartimento di Fisica, Universit\`a degli Studi di Milano, 20133 Milano, Italy}

\author{Oksana Plekan}
\affiliation{Elettra -- Sincrotrone Trieste S.C.p.A., 34149 Trieste, Italy}

\author{Kiyoshi Ueda}
\affiliation{Institute of Multidisciplinary Research for Advanced Materials, Tohoku University, Sendai 980-8577, Japan}

\author{Julian Zimmermann}
\affiliation{Institut f\"ur Optik und Atomare Physik, Technische Universit\"at Berlin, 10623 Berlin, Germany}
\affiliation{Max-Born-Institut für Nichtlineare Optik und Kurzzeitspektroskopie, 12489 Berlin, Germany}

\author{Kevin C. Prince}
\affiliation{Elettra -- Sincrotrone Trieste S.C.p.A., 34149 Trieste, Italy}
\affiliation{Department of Chemistry and Biotechnology, Swinburne University of Technology, Victoria 3122, Australia}

\author{Frank Stienkemeier}
\affiliation{Physikalisches Institut, Universit\"at Freiburg, 79104 Freiburg, Germany}

\author{Carlo Callegari}
\affiliation{Elettra -- Sincrotrone Trieste S.C.p.A., 34149 Trieste, Italy}
\affiliation{ISM-CNR, Istituto di Struttura della Materia, LD2 Unit, 34149 Trieste, Italy}

\author{Thomas Fennel}
\email[]{thomas.fennel@uni-rostock.de}
\affiliation{Institut f\"ur Physik, Universit\"at Rostock, 18051 Rostock, Germany}
\affiliation{Max-Born-Institut für Nichtlineare Optik und Kurzzeitspektroskopie, 12489 Berlin, Germany}

\author{Daniela Rupp}
\email[]{daniela.rupp@mbi-berlin.de}
\affiliation{Institut f\"ur Optik und Atomare Physik, Technische Universit\"at Berlin, 10623 Berlin, Germany}
\affiliation{Max-Born-Institut für Nichtlineare Optik und Kurzzeitspektroskopie, 12489 Berlin, Germany}

\author{Thomas M\"oller}
\affiliation{Institut f\"ur Optik und Atomare Physik, Technische Universit\"at Berlin, 10623 Berlin, Germany}


\date{December 17, 2018}

\begin{abstract}
A significant fraction of superfluid helium nanodroplets produced in a free-jet expansion has been observed to gain high angular momentum resulting in large centrifugal deformation. We measured single-shot diffraction patterns of individual rotating helium nanodroplets up to large scattering angles using intense extreme ultraviolet light pulses from the FERMI free-electron laser. Distinct asymmetric features in the wide-angle diffraction patterns enable the unique and systematic identification of the three-dimensional droplet shapes. The analysis of a large data set allows us to follow the evolution from axisymmetric oblate to triaxial prolate and two-lobed droplets. We find that the shapes of spinning superfluid helium droplets exhibit the same stages as classical rotating droplets while the previously reported metastable, oblate shapes of quantum droplets are not observed. Our three-dimensional analysis represents a valuable landmark for clarifying the interrelation between morphology and superfluidity on the nanometer scale.
\end{abstract}


\maketitle

\begin{figure*}
 \includegraphics{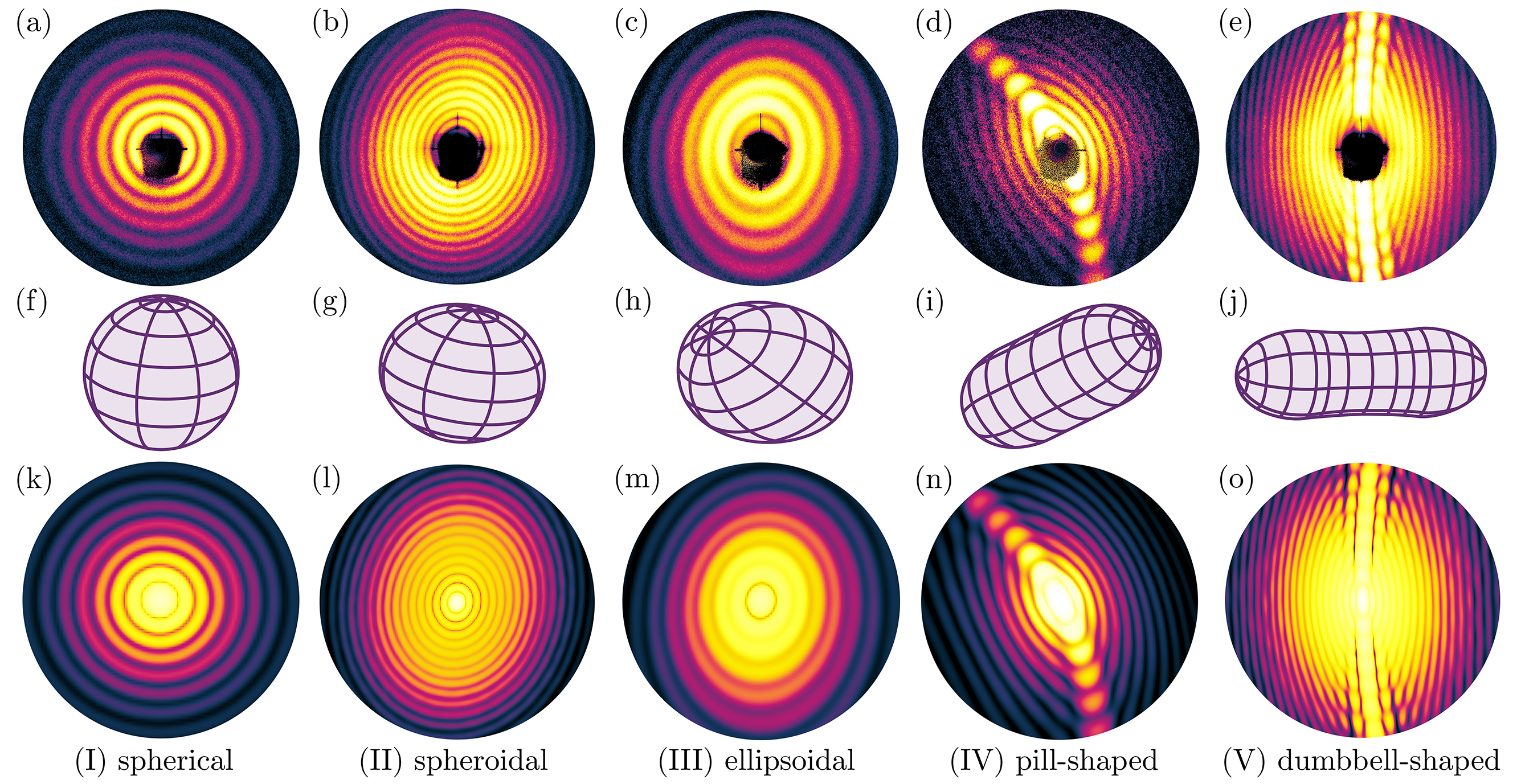}
 \caption{\label{fig:DropEvo}Evolution of spinning helium nanodroplet shapes. Experimental data (a)--(e), model shapes (f)--(j), and corresponding simulations (k)--(o); see text for details. We can classify our data into five groups (I)--(V), with a transition from spherical (f) to oblate (g) and prolate (h)--(j) shapes.}
 \end{figure*}
\begin{figure}
 \includegraphics{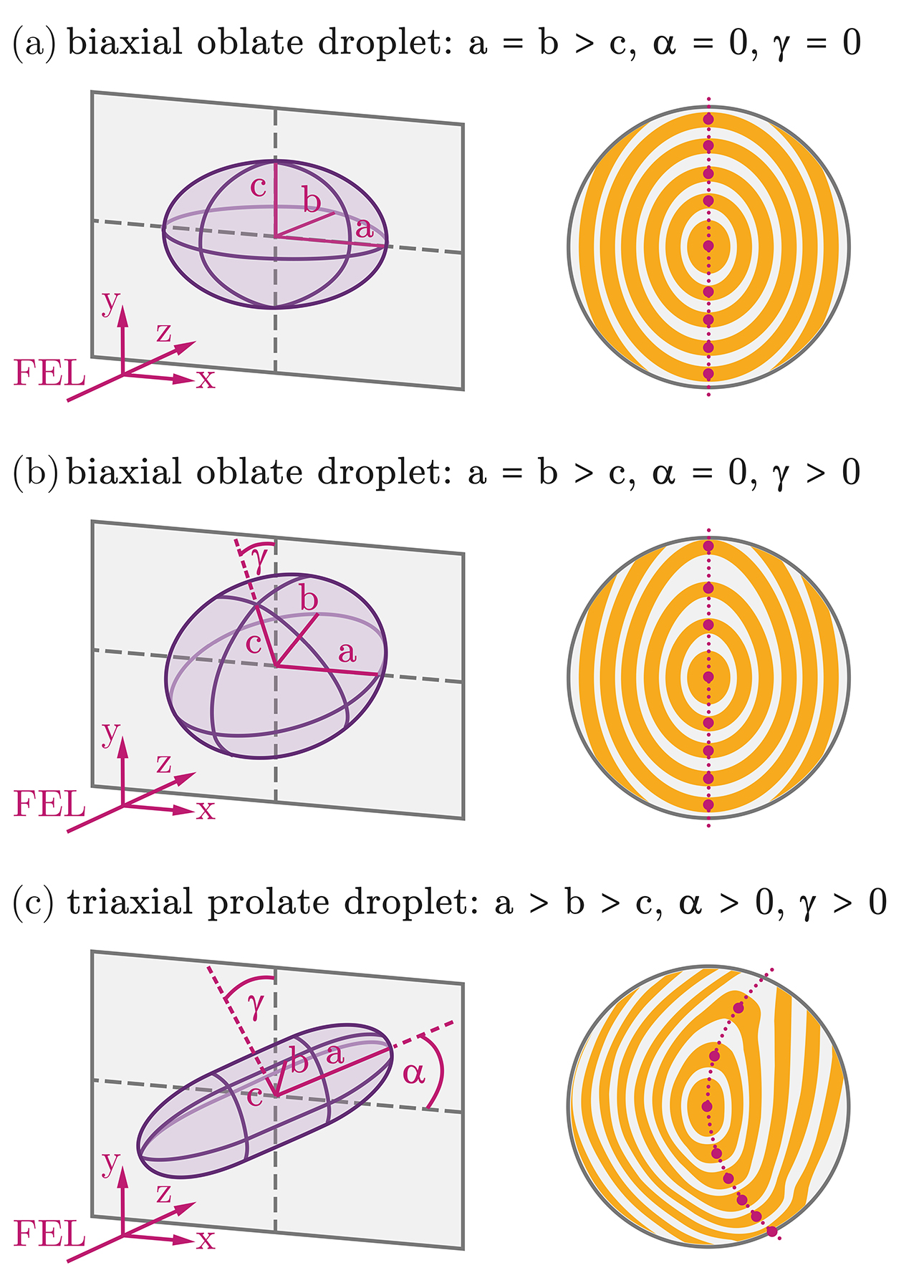}
 \caption{\label{fig:CharFeat}Origin of characteristic asymmetric features in the droplets' diffraction patterns. Any given droplet orientation can be characterized by the tilt angles $\alpha$ [rotation around $y$ axis, i.e., $\alpha = \protect\angle (\hat{x},\hat{a})$] and $\gamma$ [rotation around droplet's long principal axis, i.e., $\gamma = \protect\angle (\hat{y},\hat{c})$], where the hat denotes the unit vector of the respective axes. (a)~Oblate (biaxial) droplet, $\gamma = 0$. (b)~Oblate (biaxial) droplet. Tilting the short principal semiaxis $c$ out of the scattering plane by $\gamma > 0$ leads to a characteristic one-sided asymmetry in the diffraction pattern (indicated by a dotted line). (c)~Prolate (triaxial) droplet. Tilting the long principal semiaxis $a$ out of the scattering plane by $\alpha > 0$ leads to a bending of the diffraction pattern (see dotted line). An additional asymmetry along the bending of the scattering pattern occurs when tilting the droplet by $\gamma$ (here exemplified as two maxima versus five maxima until the detector edge). Note that rotation of the droplet around the optical axis $\hat{z}$ will only rotate the diffraction pattern.}
 \end{figure}

Inspired by the observation of the oblate deformation of planets, the first experimental study on rotating liquid drops was carried out by Plateau in 1843 \cite{Plateau1843}. Since then, the equilibrium shapes of spinning droplets \cite{Chandrasekhar1965} have been employed successfully to describe the structure and deformation of various systems ranging from atomic nuclei to astronomical objects \cite{Cohen1974}. The applicability of liquid drop models in a broad range of scientific areas has motivated, and is reflected by, extensive theoretical \cite{Brown1980, Heine2006, Baldwin2015} and experimental \cite{Wang1986, Hill2008, Baldwin2015} work and remains a subject of fundamental interest.\\
The equilibrium shape of a rotating drop from a classical liquid is generally determined by the subtle balance of surface tension and centrifugal force. At rest, only surface tension is present and in classical drops this leads to the formation of spheres. With increasing rotational frequency, the droplets' quest to realize the lowest-energy state leads to a shape evolution from an axisymmetric oblate to a triaxial prolate and a dumbbell-like two-lobed shape with the long principal axis being perpendicular to the rotational axis \cite{Cardoso2008}. 
In contrast, superfluid droplets do not rotate in the classical hydrodynamic sense as a rigid body, since the viscosity is vanishing. Spinning of superfluid droplets is characterized by an irrotational flow and vortices accommodating angular momentum \cite{Williams1974, Bauer1995}, with distinct implications for the droplets' equilibrium shapes \cite{Ancilotto2015}. 
A theoretical study comparing the shapes of superfluid droplets to those of classical droplets that was stimulated by our work has recently been published \cite{Ancilotto2018}.
Because of the lack of systematic experimental characterization methods for nanoparticles in the gas phase exhibiting statistical shape and size variations, the implications of superfluidity on the droplets' morphology have remained elusive.\\
With the advent of intense short-wavelength femtosecond light pulses from free-electron laser (FEL) facilities, the structural characterization of unsupported nanoparticles has become possible by using single-shot coherent diffractive imaging \cite{Gaffney2007, Bostedt2010, Bogan2010, Seibert2011, Chapman2011, Rupp2012, Barke2015}, including the possibility to investigate nanometer-sized superfluid helium droplets in free flight \cite{Gomez2014, Tanyag2015, Jones2016, Bernando2017}.\\
In a pioneering coherent diffractive imaging experiment at the Linac Coherent Light Source x-ray FEL, vortex arrays in rotating helium nanodroplets were made visible by doping with xenon \cite{Gomez2014}. The presence of vortices was proved and at the same time the droplets' shape projections were reconstructed from small-angle diffraction patterns. Only axisymmetric, oblate shapes were observed, with deformations up to a degree that would be unstable for classical droplets. This observation was ascribed to stability enhancement resulting from the presence of quantum vortices. It was further proposed that the transition from oblate to prolate shapes might be hindered for a superfluid droplet hosting a regular vortex array \cite{Ancilotto2015}. On the other hand, in subsequent studies using scattering techniques \cite{Bernando2017,Rupp2017}, in addition to oblate shapes, prolate \cite{Rupp2017} and two-lobed \cite{Bernando2017} droplets could be identified by modeling of the observed scattering patterns. While in Ref.~\cite{Bernando2017} the existence of classically unstable oblate shapes was further supported by a two-dimensional analysis of the power dependence of the diffraction intensity and the statistical occurrence of shape projections, scattering patterns with similar contours were interpreted in Ref.~\cite{Rupp2017} as prolate-shaped droplets. However, because of the small-angle scattering technique's \cite{Bernando2017} limitation to recover only 2D shape projections and the overall weak scattering signal and small statistics in Ref.~\cite{Rupp2017} the exact droplet dimensions could not be retrieved and the shape evolution could not be traced.\\
In this Letter we present a thorough quantitative three-dimensional characterization of the shapes of helium nanodroplets to capture the relationship between their shape and superfluidity. The combination of 3D sensitive scattering technique and a very large data set allows us to derive direct information on the droplet shape, such as all three axes $a$, $b$, $c$, and to follow the evolution of the droplets up to their stability limit. We find distinct asymmetric features in the wide-angle diffraction patterns that allow the unambiguous determination of the 3D shapes and the classification of the shapes by grouping them into five characteristic classes. Matching 3D scattering simulations to the experimental data enables a systematic comparison of our results to the shapes of classical rotating drops. Most importantly, our quantitative analysis shows that the features in the evolving shapes observed in our experiment agree with those found for classical droplets.\\
The experiment was performed at the LDM end station of the FERMI FEL-1 \cite{Allaria2012, Lyamayev2013, Svetina2015}. Helium nanodroplets were produced using a pulsed cluster source equipped with an Even-Lavie valve \cite{Pentlehner2009} that was cooled to $T_0 = 5.4 \pm 0.1\ \text{K}$ and operated with a stagnation pressure of $p_0 = 80\ \text{bar}$ resulting in a mean droplet size of $\langle N \rangle = 6 \times 10^9$ atoms per droplet (for details on the experimental setup, see Supplemental Material \cite{SuppMat}). Because of the broad size distribution ($\text{FWHM} \approx 190\ \text{nm}$), a wide range of droplet sizes is accessible in the single-shot analysis without adjustments of the expansion parameters. The droplets were irradiated with intense extreme ultraviolet pulses that were focused to a spot size below $10 \times 10\ \mu\text{m}^2$ (FWHM), yielding power densities exceeding $\text{3} \times 10^{\text{14}}\ \text{W}/\text{cm}^{2}$. To record the scattered light up to a maximum scattering angle $\Theta_{\text{max}} = 30^{\circ}$, a microchannel plate detector combined with a phosphor screen \cite{Bostedt2010} was placed $65\ \text{mm}$ away from the interaction region.\\
Our data set consists of a total of $38\,150$ bright diffraction images from $194\,500$ laser shots (overall hit rate $19.6\%$). Representative examples of the recorded diffraction patterns are shown in Figs.~\ref{fig:DropEvo}(a)--\ref{fig:DropEvo}(e).\\
Our analysis comprises three stages: (i) a qualitative analysis of the whole data set (classification of the shapes using a neural-network-based image recognition approach, see Ref. \cite{SuppMat} for details); (ii) a quantitative analysis of 20 selected patterns that allow exact determination of the droplet dimensions; (iii) the exclusion of metastable oblate shapes by analyzing all scattering images that exhibit asymmetric features.\\ 
The vast majority ($92.9\%$) of the bright scattering images exhibit concentric rings, as displayed in Fig.~\ref{fig:DropEvo}(a), while the remaining images show various pronounced deformations of the rings. Among those, we identify centrosymmetric patterns with either elliptical deformations of the rings or pronounced straight streaks, similar to the findings of previous work \cite{Gomez2014}. However, in addition to such patterns, about $2.6\%$ of our data show \emph{non}centrosymmetric features as depicted in Figs.~\ref{fig:DropEvo}(b)--\ref{fig:DropEvo}(e). These features range from directional asymmetries in the ring spacing to a pronounced bending of the pattern towards one side of the image. In particular, the asymmetry in the ring spacing can occur along one direction (e.g., from top to bottom), as exemplified in Fig.~\ref{fig:DropEvo}(b), or along two directions (e.g., from top to bottom and from left to right), as exemplified in Fig.~\ref{fig:DropEvo}(c). The latter situation is accompanied by a global curvature of the pattern towards one side that becomes most striking in the case of the strongly bent streaks [Figs.~\ref{fig:DropEvo}(d) and \ref{fig:DropEvo}(e)].
Earlier work has shown that such deviations from point symmetry are a clear indication of 3D information being encoded in the wide-angle diffraction patterns \cite{Barke2015}: While small-angle scattering patterns only contain 2D structural information (density projected on a \emph{single} plane normal to the optical axis), wide-angle scattering provides true 3D information as \emph{multiple} projection planes contribute \cite{Barke2015}.\\
In the following we establish the relation between particular shapes and the corresponding asymmetries in the diffraction patterns (Fig.~\ref{fig:CharFeat}). We would like to emphasize that the relevant characteristic features allow for a clear distinction between oblate and prolate droplet shapes.
In general, the spacing of the rings in the diffraction pattern is determined by the lengths of the droplet's principal semiaxes $a$, $b$ and $c$; the longer the axes, the smaller the spacing. A spherical droplet with all axes being equal will lead to concentric circles in the diffraction pattern. For an oblate droplet (i.e., axisymmetric or biaxial, $a = b > c$) that is oriented with its short axis perpendicular to the FEL axis, the diffraction pattern shows concentric elliptical rings; see Fig.~\ref{fig:CharFeat}(a). Rotating an oblate droplet out of the scattering plane, i.e., tilting the short principal semiaxis $c$ by an angle $\gamma$, will lead to a noncentrosymmetric diffraction pattern [Fig.~\ref{fig:CharFeat}(b)]. In particular, the wider fringes in the upper part of the image result from a projection that dominantly reflects the short principal axis $c$. In turn, the narrower fringes in the lower part of the image describe the longer principal semiaxis $b$. Consequently, the deformation and orientation of the particle are encoded in the ring spacing within the image [see dotted line in Fig.~\ref{fig:CharFeat}(b)]. The asymmetry of the ring spacing is maximal for $\gamma = 45^{\circ}$. Finally, Fig.~\ref{fig:CharFeat}(c) shows a triaxial prolate droplet ($a > b > c$). Tilting the long principal semiaxis $a$ out of the scattering plane by an angle $\alpha$ leads to a distinct crescentlike distortion of the elliptic ring features to one side of the scattering image [cf. the dotted line in Fig.~\ref{fig:CharFeat}(c) as a guide to the eye]. When further increasing the length of the principal semiaxis $a$, the diffraction pattern will eventually show the pronounced streaks which are also bent given that $\alpha > 0$.\\
The above considerations demonstrate that the asymmetric features can be understood and associated with distinct shape parameters. However, in contrast to the case of hard x-ray small-angle scattering, where the 2D projection of the particle's electron density distribution can be uniquely reconstructed from the diffraction pattern via iterative phase retrieval algorithms \cite{Marchesini2003, Seibert2011, Tanyag2015}, there is so far no rigorous reconstruction algorithm available for wide-angle scattering. We therefore employ a simple parametrized shape model for the droplets and compute wide-angle diffraction patterns that are iteratively matched to the recorded images in a forward-fitting procedure.
In particular, we consider a structure consisting of two ellipsoidal caps and a hyperboloidal centerpiece, controlled by five input parameters, and calculate the corresponding wide-angle diffraction patterns with a generalized version of the multislice Fourier transform algorithm in Ref.~\cite{Barke2015}.
The accuracy of the fit-based shape retrieval is limited by the information content and quality of the scattering image data (e.g., due to the maximal recorded scattering angle $\Theta_{\text{max}} = 30^{\circ}$, an inhomogeneous detector sensitivity, and detector saturation effects).  As particle tilt angles near $\gamma = 45^{\circ}$ lead to the strongest directional asymmetries in the fringe spacing and thus to the highest sensitivity of the retrieval procedure for droplet deformation, we selected these optimally oriented droplets for our analysis (in total 20 patterns).
The results of the shape analysis are shown for a selection of representative images in Fig.~\ref{fig:DropEvo}; see Supplemental Material for technical details \cite{SuppMat}. The images reflect spherical [Fig.~\ref{fig:DropEvo}(f)], oblate [Fig.~\ref{fig:DropEvo}(g)], as well as prolate [Figs.~\ref{fig:DropEvo}(h)--\ref{fig:DropEvo}(j)] shapes. 
Based on asymmetry features, we have defined five shape groups as indicated in Fig.~\ref{fig:DropEvo} and classified our whole data set using the neural network approach described in the Supplemental Material \cite{SuppMat}. The results of our classification analysis yield the following shape distribution: (I) spherical (concentric circles, $92.9\%$), (II) spheroidal (elliptical patterns or one-sided asymmetry, $5.6\%$), (III) ellipsoidal (bent patterns, $0.8\%$), (IV) pill-shaped (streaked patterns, $0.6\%$), and (V) dumbbell-shaped (streaks with side maxima or pronounced side minima, less than $0.1\%$).\\
 \begin{figure}
 \includegraphics{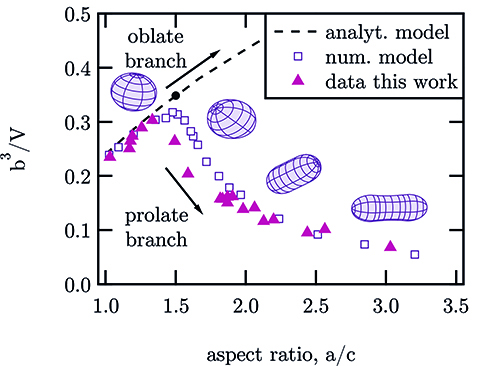}
 \caption{\label{fig:AR}Ratios of the principal semiaxis lengths $a$, $b$, $c$, and the droplets' volume $V$. The dimensionless ratio $b^3/V$ is plotted versus the aspect ratio $a/c$. Dashed line: Analytical model for axisymmetric equilibrium shapes of rotating droplets \cite{Chandrasekhar1965}. Squares: Numerical model for classical droplet shapes of spinning droplets \cite{Baldwin2015}. Triangles: This experiment. Our data follow the oblate (axisymmetric) branch up to an aspect ratio of $a/c\approx 1.5$ and then evolve along the prolate (triaxial) branch, with pill-shaped droplets starting at $a/c>1.8$ and dumbbell-like shapes when $a/c>2.5$. The stability limit for classical axisymmetric droplets is indicated by the filled dot. To visualize the droplet shape evolution, model shapes from Fig. \ref{fig:DropEvo} are reproduced.}
 \end{figure}
For comparison of our experimental data to the prediction of the numerical model for classical droplet shapes with varying angular momentum proposed by Baldwin \emph{et}~al.~\cite{Baldwin2015}, we plot both data sets using the same dimensionless ratios of the shape parameters; see squares and triangles in Fig.~\ref{fig:AR}. Please note that droplets with the same aspect ratio do not possess equal angular momentum in normal and superfluid states. The helium droplet shapes follow first the oblate branch predicted by an analytical model for axisymmetric droplets~\cite{Chandrasekhar1965} up to an aspect ratio $a/c$ of about $1.5$ and then continue to evolve along the prolate (triaxial) branch with only slight deviations from the numerical model for classical liquids \cite{Baldwin2015}. For aspect ratios $a/c>1.8$, the droplets become pill shaped and for $a/c>2.5$, they exhibit a dumbbell-like shape. Most importantly, no axisymmetric droplet shapes have been observed beyond the classical limit of instability for oblate drops \cite{Chandrasekhar1965} marked by the filled black circle in Fig.~\ref{fig:AR}. Up to aspect ratios as large as $a/c = 3.0$, our data show classical behavior. By classifying all diffraction patterns we ensured that the key findings of our quantitative shape reconstruction are valid for the whole data set: (i) We identified all images exhibiting strong asymmetries that are only expected for extremely deformed shapes. (ii) No indication for the characteristic diffraction patterns of metastable droplet shapes (see Supplemental Material \cite{SuppMat}) was found. (iii) We could trace the transition from biaxial oblate to triaxial prolate and dumbbell-like shapes.
Our results can be further compared with recent theory work on the shapes of spinning superfluid droplets~\cite{Ancilotto2018}. Two classes of droplet shapes were predicted, one containing vortices and a second class of vortex-free configurations. For aspect ratios $a/c>1.5$, Ancilotto and co-workers \cite{Ancilotto2018} find that the shapes of both classes resemble the classical droplet model, but for $a/c<1.5$ (i.e., low angular momenta), the vortex-free superfluid droplets differ from the classical model. Our data are in good agreement with classical shapes and therefore with the shapes of droplets containing vortices. These results suggest that the presence of vortices is the reason why the shapes of superfluid spinning droplets follow those of classical droplets~\cite{Ancilotto2018}.\\
The results of our analysis and the recent theory work point to fundamental questions on the nature and evolution of the normal liquid to superfluid phase transition of helium nanodroplets during formation. When expanded from the normal liquid phase, one could imagine a nonsuperfluid helium nanodroplet to first gain rotational energy, following the equilibrium shape of a classically rotating drop, and then undergo the transition to a superfluid as soon as the droplet temperature drops below the $\lambda$ line at a later stage of the liquid jet expansion. The exact time of the phase boundary crossing may depend on the nozzle geometry, pressure and temperature of the helium, and maybe even on the type of source used (continuous or pulsed). For example, a 3D sensitive experiment with the source of Ref.~\cite{Bernando2017} would be needed to verify if the occurrence of metastable shapes is related to the droplet preparation process.
We assume that the transition to the superfluid phase triggers the formation of vortices. Because of conservation of angular momentum, the phase transition is presumably accompanied by a change of droplet shape \cite{Ancilotto2018}. Therefore, it would be very interesting to further investigate the dynamics of the phase transition. Using the new two-color capabilities at FEL facilities \cite{Ferrari2016}, one could imagine simultaneously recording the 3D shape with a long wavelength (extreme ultraviolet) pulse as well as the vortex array structure of a prolate superfluid droplet with a short-wavelength (x-ray) pulse.\\
To summarize, we have observed the shape deformations of spinning helium nanodroplets and followed the complete evolution from oblate to prolate shapes by exploiting the 3D structure information contained in single-shot wide-angle scattering data. 
All observed droplet geometries are in surprisingly good agreement with the shapes of classically rotating droplets obtained from numerical simulations \cite{Baldwin2015} with only small deviations in the transition regime from oblate to prolate shapes (i.e., $a/c \approx 1.5$). These small deviations could be an indication for different vortex configurations inside the droplets \cite{Ancilotto2018}. While the underlying physics should be clarified in follow-up work, we note that the qualitative trend of our data is fully compatible with the classical picture and in strong contrast to previously reported metastable oblate shapes \cite{Gomez2014, Bernando2017}. The proposed metastable ``wheel-shaped'' quantum droplets \cite{Gomez2014} are not observed. Further, the shapes are in good agreement with recently predicted shapes for spinning superfluid droplets containing vortices \cite{Ancilotto2018}. Finally, for small angular momenta, we have presently no evidence for the realization of the shapes predicted for droplets that do not contain vortices \cite{Ancilotto2018}. 
We believe that our results stimulate the presently very active field of research on strongly correlated, liquidlike quantum systems, which cover an enormous range of densities from Bose-Einstein condensed particles, as in recently suggested dark matter superfluidity \cite{Berezhiani2015, Berezhiani2018}, to dipolar gases \cite{Schmitt2016}, liquid drops and nuclear matter, as in neutron stars \cite{Lattimer2004,Jones2010}.

\begin{acknowledgments}
We would like to thank A. Vilesov, F. Ancilotto, and M. Barranco for fruitful discussions and acknowledge excellent support of the FERMI staff during the beam time. This work received financial support by the Deutsche Forschungsgemeinschaft under the grant MO 719/14-1, by the Bundesministerium f\"ur Bildung und Forschung (BMBF, project ID 05K16KT3), and by the Leibniz grant SAW/2017/MBI4. T.F. acknowledges financial support from the Deutsche Forschungsgemeinschaft via SFB 652, via a Heisenberg fellowship (No. FE 1120/4-1), and from the Bundesministerium f\"ur Bildung und Forschung (BMBF, Project No. 05K16HRB). Computing time has been provided by the North German Supercomputing Alliance (HLRN, No. mvp00013). M.D. acknowledges funding by the Swiss National Science Foundation through Grants No. 200021\_146598 and No. 200020\_162434. P.P. acknowledges support from Italian Ministry of Education, Universities and Research (MIUR) PRIN 2012Z3N9R9 ``NOXSS.'' F.S. acknowledges support by the Deutsche Forschungsgemeinschaft under the Grant No. STI 125/19-1. K.U. acknowledges support by the XFEL Priority Strategy Program MEXT and the TAGEN project while T.N. and K.U. acknowledge support by the Research Program of ``Dynamic Alliance for Open Innovation Bridging Human, Environment and Materials'' in ``Network Joint Research Center for Materials and Devices.''
\end{acknowledgments}


%

\end{document}